\documentclass[aps,prl,preprint,groupedaddress,showpacs]{revtex4}
\usepackage{graphicx}
\usepackage{amsmath}
\begin{document}

\title{Gate-tuned high frequency response of carbon nanotube Josephson junctions}

\author{J.-P. Cleuziou$^1$, W. Wernsdorfer$^2$, S. Andergassen$^2$, 
S. Florens$^2$, V. Bouchiat$^2$, Th. Ondarcuhu$^1$, M. Monthioux$^1$}

\affiliation{
$^1$CEMES-CNRS, BP 94347, 31055 Toulouse Cedex 4, France\\
$^2$Institut N\'eel, CNRS/UJF, BP 166, 38042 Grenoble Cedex 9, France
}

\date{April 20, 2007}

\begin{abstract}
Carbon nanotube (CNT) Josephson junctions in the open quantum
dot limit exhibit superconducting switching currents which can 
be controlled with a gate electrode. 
Shapiro voltage steps can be observed under radiofrequency
current excitations, with a damping of the phase dynamics that strongly 
depends on the gate voltage.
These measurements are described by a standard RCSJ
model showing that the switching currents 
from the superconducting to the normal state
are close to the critical current of the junction.
The effective dynamical capacitance of the nanotube junction is 
found to be strongly gate-dependent, suggesting a diffusive 
contact of the nanotube.
\end{abstract}


\maketitle

It was recently demonstrated that carbon nanotube (CNT) 
weak links, connecting two superconducting leads~\cite{Kasumov99}, 
can be gate-controlled Josephson 
junctions~\cite{Buitelaar02,Buitelaar03,Jarillo_Herrero06,Jorgensen06,Cleuziou_NatureNano06,Tsuneta07,Eichler07} 
and even gate-controlled $\pi$-junctions~\cite{Cleuziou_NatureNano06}. 
The first interesting application of CNT Josephson junctions are
superconducting transistors~\cite{Jarillo_Herrero06,Jorgensen06} and 
superconducting quantum interference devices
(SQUIDs)~\cite{Cleuziou_NatureNano06}, which are promising candidates for the detection 
of individual magnetic molecules. 
They might also prove useful as building blocks for more complicated 
superconducting devices~\cite{ChiorescuScience03} and new readout 
schemes~\cite{Siddiqi04,Siddiqi05}. 

For all theses devices, the critical current $I_{\rm c}$ and the high 
frequency response of CNT Josephson junctions plays an important role, which is 
studied in this letter. Indeed, previous investigations showed that
the magnitude of the switching current $I_{\rm sw}$ of such CNT
devices appeared to be very different from theoretical predictions: in the 
first experiments it was observed to be 10 times larger~\cite{Kasumov99} 
and later 10 times 
smaller~\cite{Jarillo_Herrero06,Jorgensen06,Cleuziou_NatureNano06,Tsuneta07,Eichler07} 
than the result of Ambegaokar and Baratoff~\cite{Ambegaokar_Baratoff63},  
reformulated for a quantum dot (QD) in ~\cite{Beenakker92}. Moreover, all these previous 
measurements were performed for a static bias, without retrieving the effect of
radiofrequency (RF)  irradiation.

The electronic transport through a CNT junction 
can be described by a QD in between 
two conducting leads. The energy level spacing $\Delta E$
depends on the length of the junction and the 
nature of the conducting leads. The transport can be 
classified into three limits depending on the ratio 
between the transparency $\Gamma$ of the QD barriers and 
the charging energy $U$~\cite{note1}: (i) for $h\Gamma << U$, the maximum 
conductance at zero bias obeys $G_{\rm max} << 2e^2/h$. In this so-called 
closed QD regime the charging effects 
dominate the transport characterized by well-resolved Coulomb 
blockade diamonds~\cite{Cobden02}. (ii) for $h\Gamma \approx U$
holds $e^2/h \lesssim G_{\rm max} \lesssim 2e^2/h$. In this
intermediate transparency regime charging effects as well as 
higher-order tunneling processes are very important. 
Here, transport measurements show that Coulomb blockade 
diamonds, corresponding to an odd number of electrons, are connected 
by Kondo ridges~\cite{Goldhaber_Gordon98,Buitelaar02}. 
For superconducting leads, higher-order 
multiple Andreev reflections (MARs) are well 
resolved~\cite{Buitelaar03,Jarillo_Herrero06}. 
(iii) For $h\Gamma >> U$, the conductance is close to $4e^2/h$ 
and corresponds to the open QD regime
(the factor 4 reflects 
the two modes of the CNT band structure as well 
as the two spin orientations), 
where Coulomb blockade diamonds 
are not observed. This leads to a relatively high 
supercurrent for all gate voltages.
Gate-modulated Fabry-Perot interferences can be observed 
in the last two regimes~\cite{Liang01}. 

An important advantage of these devices is that backgate and sidegate electrodes 
can be used to vary both $\Gamma$ and the energy levels position in the nanotube, 
thus permitting to change the electric transport regime~\cite{Cleuziou_condmat06}.
This tunability allowed us in a previous study to capture both Kondo screened
and unscreened junctions, leading to the observation of a $0$-$\pi$ transition~\cite{Cleuziou_NatureNano06}. 
Similarly, we found that the QD barrier can be varied with the thickness of the
Pd contacts, enabling to access also all three transport regimes.

Previous measurements on CNT Josephson junctions 
were in the intermediated or closed QD regime~\cite{note2}. 
Here we present the first measurements in the 
open QD regime using a Pd thickness of 7 nm. The present description
concentrates on a single device, similar data 
were however obtained on a couple of other devices fabricated 
on the same chip.

The CNT Josephson junction were fabricated as presented 
earlier~\cite{Cleuziou_NatureNano06}. We started from a degenerately 
n-doped silicon substrate with 350 nm thick thermally 
grown SiO$_{2}$ layer on top used as a backgate. 
Single-walled CNTs were deposited using a combing 
technique~\cite{Gerdes99}. 
The nanotube location was imaged by atomic force microscopy (AFM) 
and aligned e-beam lithography patterned
the contacts~\cite{Sagnes03}. The length of the tube section between the 
contacts was about 200 nm. 
Metal electrodes consisting of a thickness of 3 to 7 nm of Pd followed 
by 50 nm of Al, were deposited using electron-gun evaporation.
Pd provides high-transparency 
contacts to the carbon nanotubes. Al is a superconductor 
having a critical temperature of about 1.2 K. 
The room temperature resistance 
varied between 10 and 60 k$\Omega$, strongly depending on the 
thickness of Pd. Only devices with no significant gate 
effect at room temperature were used for our studies. 
We found that for Pd thicknesses below about 3 nm, the 
CNT junction was in the closed QD regime, for 4 to 6 nm 
in the intermediate, and for 7 nm and more in the open QD regime.

The measurements were performed in a shielded cryostat 
having heavily filtered lines in order to minimize 
the electronic noise reaching the sample, since 
electronic noise can reduce and even suppress the 
switching current of the CNT-SQUID. Our home-built filtering 
system was presented in~\cite{Cleuziou_NatureNano06}. We used mainly a 
four-probe, current-biased method.
with a bias resistance 
of about 65 M$\Omega$ at 4.5 K and voltage dividers 
at room temperature. The voltage signal of the sample was 
first pre-amplified by a factor of 10$^4$ and then measured 
using an oscilloscope or a look-in amplifier. 
High frequency ac current modulations in the 
frequency range between 1 and 12 GHz were induced via an 
inductive coupling.

The effective BCS gap $\Delta_{\rm eff}$ of the superconducting 
leads felt by the CNT was estimated by voltage versus 
current measurements as a function of temperature. 
We found a transition temperature $T_{\rm c}$ = 0.7 K, 
yielding a $\Delta_{\rm eff}$ = 1.76 $k_{\rm B}T_{\rm c}$ = 0.1 meV. 
The energy level spacing of the CNT were determined 
$\sim 10$ meV~\cite{Cleuziou_NatureNano06}. 
However, due to the high transparency of the contacts the effect of the Coulomb 
energy is much smaller here.

\begin{figure}
\begin{center}
\includegraphics[width=.5\textwidth]{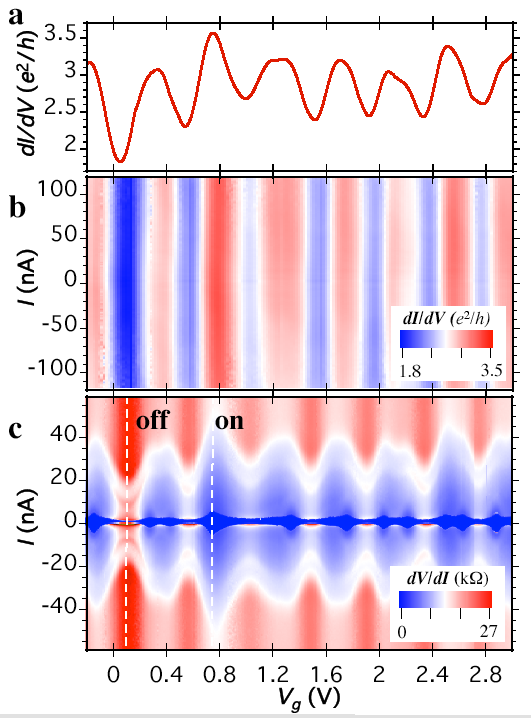}
\caption{(Color online) Electric transport through a CNT junction as a function 
of gate-voltage $V_{\rm g}$. (a) Differential 
conductivity $dI/dV$ versus $V_{\rm g}$ at zero bias. 
(b) $dI/dV$-map versus current bias $I_{\rm sd}$ and 
$V_{\rm g}$ in the normal state ($H_z$ = 50 mT). 
(c) Differential resistivity $dV/dI$-map versus $I_{\rm sd}$ and 
$V_{\rm g}$ in the superconducting state ($H_z$ = 0). 
The dotted lines indicates the off-state 
and on-state, which are further studied in Figs. 2 to 4.
}
\label{fig1}
\end{center}
\end{figure}

In Fig. 1a and 1b, a magnetic field of $H_z$ = 50 mT was 
applied perpendicular to the electrode plane in order 
to suppress the superconductivity of the leads. 
The differential conductivity $dI/dV$ is plotted 
versus gate voltage at zero bias in Fig. 1a and, in addition, 
as a function of source-drain current $I_{\rm sd}$ in 
Fig. 1b. $dI/dV$ oscillates between about 1.8 and 3.5 $e^2/h$ 
and is $I_{\rm sd}$-independent in the considered range. $dI/dV$ 
is maximal when a quantum level of the CNT is aligned with respect 
to the Fermi energy of the leads. A current can then 
flow by resonant tunneling through the CNT, and we denominate
this regime as an on-state. When the quantum levels are 
far from the Fermi energy, the current is reduced (off-state). 
This CNT junction is clearly in the open QD regime because the 
reduction is only about a factor of two and Coulomb blockade 
effects are not directly observable.
 
Fig. 1c presents the same measurements as in Fig. 1b 
but with superconducting leads ($H_z$ = 0). Because $dI/dV$ diverges, 
the differential resistivity $dV/dI$ is plotted. For small $I_{\rm sd}$ 
a supercurrent is observed, which is evidenced by a nearly 
zero-resistance state. At a certain bias current 
denoted by switching current  $I_{\rm sw}$, the junction 
switches from the superconducting to the normal state.  $I_{\rm sw}$ 
is strongly gate-voltage dependent: it is maximal (minimal) 
in the on-state (off-state). Most of the on-states are 
split into two  $I_{\rm sw}$ maxima suggesting 
that there is still a small influence of Coulomb interaction. 
Note that higher-order MARs are not observed, which is 
probably due to strong broadening effects induced by the 
high transparency contacts.

\begin{figure}
\begin{center}
\includegraphics[width=.55\textwidth]{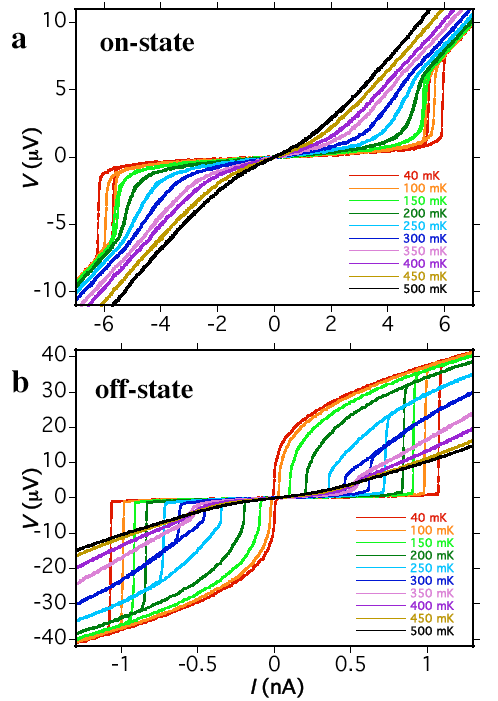}
\caption{(Color online) Voltage versus current characteristics for a CNT 
junction in (a) the on-state and (b) off-state at several 
temperatures. (a) and (b) correspond to maximal and minimal 
switching currents in Fig. 1c, respectively. The current 
was swept at the sweep rates of 5 and 0.2 nA/s for (a) 
and (b), respectively. Note that each curve is a single 
sweep (no data averaging was performed) and the slope 
at the switching currents was limited by the filtering.}
\label{fig2}
\end{center}
\end{figure}

Fig. 2a and 2b show typical voltage versus current curves 
at several temperatures for the on- and off-state. 
The on-state is characterized by the largest $I_{\rm sw}$ and 
small hysteresis effects, that is the retrapping current 
is close to $I_{\rm sw}$. Furthermore, $dV/dI$ increase for 
$I_{\rm sd} > I_{\rm sw}$ in the considered range. 
In the off-state, the hysteresis effects are very large and $dV/dI$
decrease for $I_{\rm sd} > I_{\rm sw}$.
A small temperature-dependent resistance is observed at zero-bias,
which is mainly due to temperature-induced phase 
diffusion~\cite{Vion96}. 

The high frequency response of the CNT junctions is examined by 
voltage-current curves in the presence of RF-fields. Figs. 3a and 4a 
show the appearance of constant voltage Shapiro steps, 
which are due to the harmonical synchronization of 
the Josephson oscillations and the applied RF excitations. 
The steps appear at voltages equal to $n f \phi_0$ 
where $n = 1, 2, ...$  and $\phi_0$ is a flux quanta. 
We checked this linear relation in the frequency 
range between $f$ = 1 to 15 GHz (data not shown). 
The influence of the RF current $I_{\rm RF}$ on the Shapiro steps is 
presented on Figs. 3b and 4b for the on- and off-state,  respectively.
The plots show the differential resistance 
as a function of $I_{\rm RF}$ and $I_{\rm sd}$, giving rise to a complicated 
pattern of non-dissipative lobes.

We modelled the data with a phenomenological description of time-dependent 
transport in superconducting junctions based on both the standard Josephson 
relations and on classical circuit theory, the so-called resistively and capacitively 
shunted junction (RCSJ) model~\cite{Barone82,Tinkham96}. This invokes simple differential 
equations that govern the dynamics of the phase $\phi$ across the junction:
\begin{subequations}
\begin{eqnarray}
\hbar \frac{d\phi}{d t} & = & 2 e V \\
I(t) & = & C \frac{d V}{d t} + I_{\rm qp}(V) + I_{\rm c} \sin\phi \;,
\end{eqnarray}
\end{subequations}
where $V$ is the instantaneous voltage drop, 
$I(t)=I_{\rm sd}+I_{\rm RF}\sin(\omega t)$
the injected current with an oscillating part at the radiofrequency 
$f=\omega / 2\pi$,
$I_{\rm qp}(V)$ the quasiparticle contribution to the current, $I_{\rm c}$ the 
critical supercurrent, and $C$ the capacitance of the junction.

Assuming an ohmic quasiparticle current $I_{\rm qp}=V/R$ and defining a 
basic frequency scale $\gamma=2eRI_{\rm c}/\hbar$, one obtains two relevant 
dimensionless parameters in the analysis of the Shapiro 
steps: i) the reduced pulsation $\tilde\omega=\omega/\gamma$ characterizing
the shape of the Shapiro steps~\cite{Likharev71,Russer72} (pure Bessel functions 
are obtained in the limit $\tilde\omega\gg1$ only); ii) the quality factor 
$Q=\sqrt{\gamma RC}$ that controls the damping of the phase dynamics. 

The parameters are adjusted to the experimental data in the 
RF driven regime where the dynamics is less sensitive to phase
diffusion~\cite{Vion96}.
The analysis of the experimental data in the on-state reveals a globally good 
agreement for $Q \lesssim 1$, i.e. the  RSJ model with no capacitance.
Indeed, the geometrical capacitance of CNT junctions is very small 
($\sim 30$ aF). This finding is supported by the small hysteresis
effects of the $VI$-curves (Fig. 2a). Note that the predicted $I_{\rm 
c}$ is very close to $I_{\rm sw}$ at $I_{\rm RF}$ = 0.
The Shapiro step positions present however small but
appreciable deviations to the pure Bessel functions. Their precise shape can
indeed be more accurately described using a reduced pulsation
$\tilde\omega\sim 0.35$. 
Furthermore, a clear downward dispersion of the lobes is observed, which
may be attributed to the non-ohmic character of the quasiparticle current
$I_{\rm qp}(V)$ above the threshold $I_{\rm sw}$. 
Simulations of the phase dynamics with such non-linear RSJ model, where 
the non-linear $R$ from the $VI$-curves in the resistive 
state of Fig. 2 was included, are provided in Fig. 3c and compare 
favorably to the experimental data.

\begin{figure}
\begin{center}
\includegraphics[width=.52\textwidth]{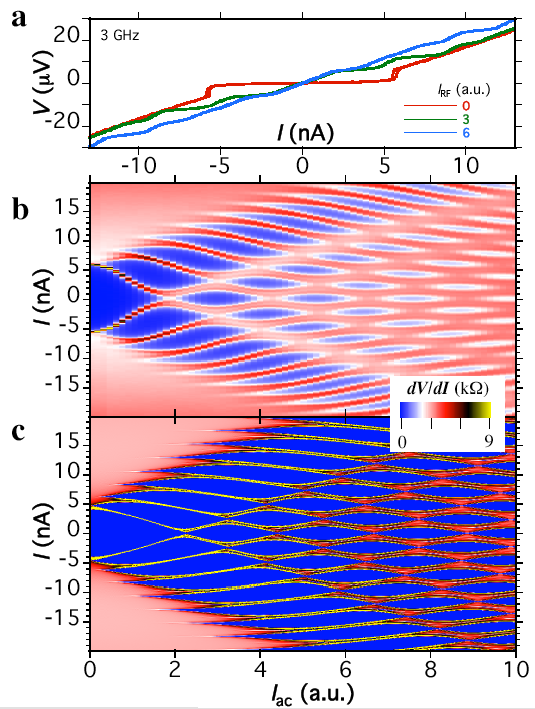}
\caption{(Color online) RF response in the on-state.
(a) Voltage versus current characteristics at several RF amplitudes
$I_{\rm RF}$. (b) Differential resistance $dI/dV$-maps versus 
$I_{\rm sd}$ and $I_{\rm RF}$ for $f$ = 3 GHz. (c) Simulation of the data in (b)
using the non-linear RSJ model.
}
\label{fig3}
\end{center}
\end{figure}

\begin{figure}
\begin{center}
\includegraphics[width=.52\textwidth]{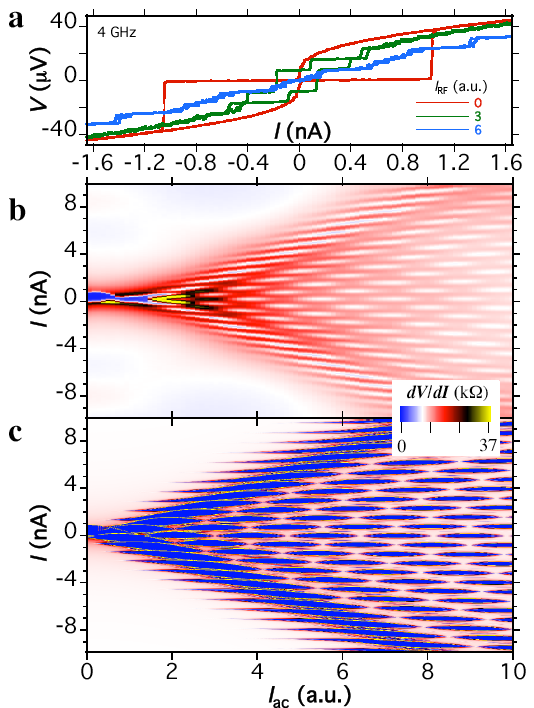}
\caption{(Color online) RF response in the off-state.
(a) $VI$-characteristics at several $I_{\rm RF}$. 
(b) $dI/dV$-maps versus 
$I_{\rm sd}$ and $I_{\rm RF}$ for $f$ = 4 GHz. (c) Simulation of the data in (b)
using the non-linear RCSJ model.
}
\label{fig4}
\end{center}
\end{figure}

Regarding the off-state, several issues point to additional underdamping effects:
the $VI$-curves are strongly hysteretic and the degree of
hysteresis diminishes with increasing temperature (Fig. 2b)~\cite{Tinkham96}. Moreover,
the low-bias Shapiro steps present a peculiar threshold behavior 
(Fig. 4b), while the lobes show an upward dispersion.
These properties 
can not be explained within the usual RCSJ model, and may be only accounted for 
at a qualitative level by introducing a rather large effective capacitance
$C \approx 1 $ fF and a non-linear quasiparticle resistance (stronger than in 
the on-state discussed above). Fig. 4c shows a partial agreement between our simulations 
and the data. The gate-dependence of $C$ suggests a diffusive CNT 
contact in the off-state, i.e. additional capacities coming from the CNT 
contacts. Indeed, for the off state, the electrode/nanotube capacitance
has to be taken into account 
since the Pd/nanotube contact is known to be 
distributed over a larger part of the contact~\cite{Nemec06}.
Its value can be rather strong ($\sim$ 1 fF). For the on-state however, the capacitance 
of the contact  can be strongly renormalized by the tunnel effect~\cite{Flensberg93} 
and the capacitance of 
the junction is mainly due to the nanotube self-capacitance 
estimated as 30 aF for a nanotube portion of 200 nm~\cite{note3}. 
A more thorough analysis of quantum and thermal fluctuation effects on the
Shapiro steps in the spirit of references~\cite{Ivanchenko69,Duprat05} is clearly 
beyond the scope of the present study. We point out however that these 
effects probably do not account for the strong reduction of the 
critical current, as the above analysis demonstrated in the on-state.

In conclusion, we studied the critical currents and the constant-voltage
Shapiro steps under RF irradiation of a strongly coupled CNT
Josephson junction, and discovered a strong gate-voltage dependence.
These junctions are therefore very promising
as tunable RF building blocks for superconducting devices.

This work was partially supported by the CNRS ACI NOCIEL 
and ANR QuSpins programs. Clean room processes 
were supported by the CNRS RTB program using the 
technological facilities of the Laboratory for Analysis 
and Architecture of Systems (University of Toulouse - France). 
We thanks F. Balestro, O. Buisson, F. Carcenac, J. Clarke, 
E. Eyraud, D. Feinberg, and I. Siddiqi for valuable contributions and helful discussions.


\begin{thebibliography}{10}
\bibitem{note2}
The data in~\cite{Jarillo_Herrero06} were interpreted to be 
in the open QD regime. However, we think that the 
well-resolved higher-order MAR fine-structures as well 
as the high resistivity in the off-states suggest 
that these devices operated in the intermediate transport regime. 
The observation of $G_{\rm max} \approx 4e^2/h$ 
could be explained by the large CNT diameter of 4 to 7 nm
and the observation of Fano resonances suggesting that the 
junction was not made out of an individual single-walled CNT.
\end{thebibliography}

\section{Annex}
\begin{figure}
\begin{center}
\includegraphics[width=.65\textwidth]{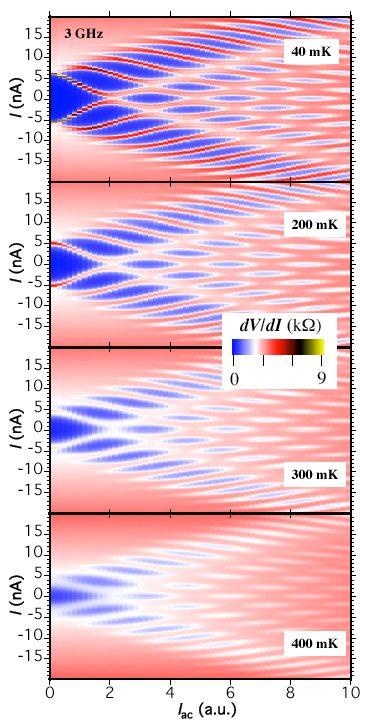}
\caption{(Color online) RF response in the on-state
at four different temperatures.
$dI/dV$-maps versus 
$I_{\rm sd}$ and $I_{\rm RF}$ are displayed for $f$ = 3 GHz. }
\label{figA1}
\end{center}
\end{figure}

We present here additional data 
measured at different temperatures and microwave frequencies.
Fig.~\ref{figA1} shows differential resistance maps for the on-state
at several temperature. Although the Shapiro steps are blurred because
of phase diffusion induced by the temperature, the step positions
are nearly temperature independent.

\begin{figure}
\begin{center}
\includegraphics[width=.49\textwidth]{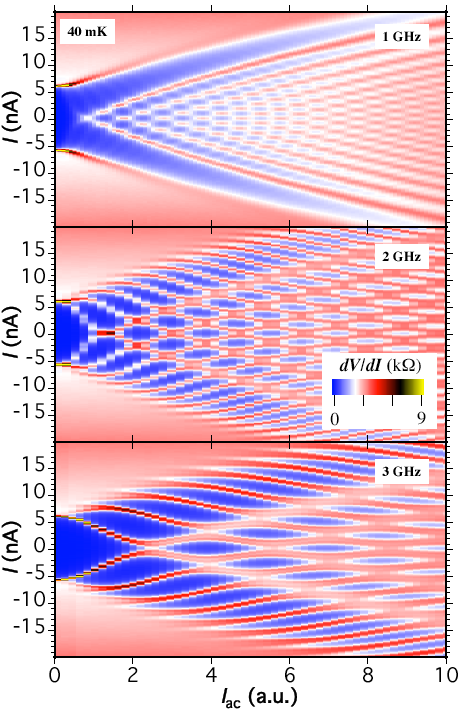}
\includegraphics[width=.49\textwidth]{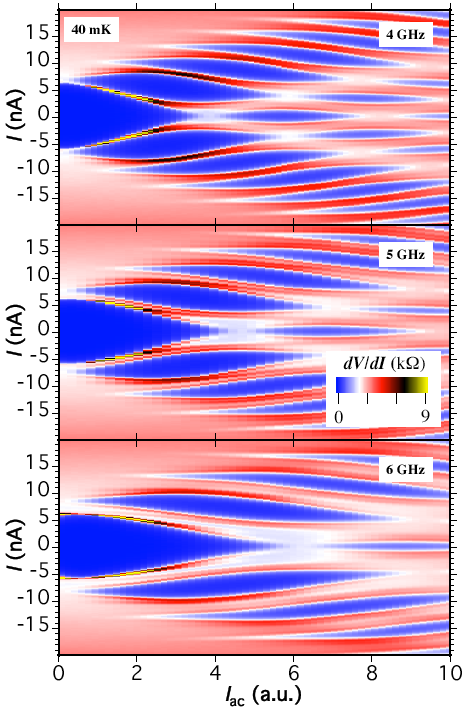}
\caption{(Color online) RF response in the on-state
at different microwave frequencies.
$dI/dV$-maps versus 
$I_{\rm sd}$ and $I_{\rm RF}$ are displayed for 40 mK. }
\label{figA2}
\end{center}
\end{figure}

\begin{figure}
\begin{center}
\includegraphics[width=.8\textwidth]{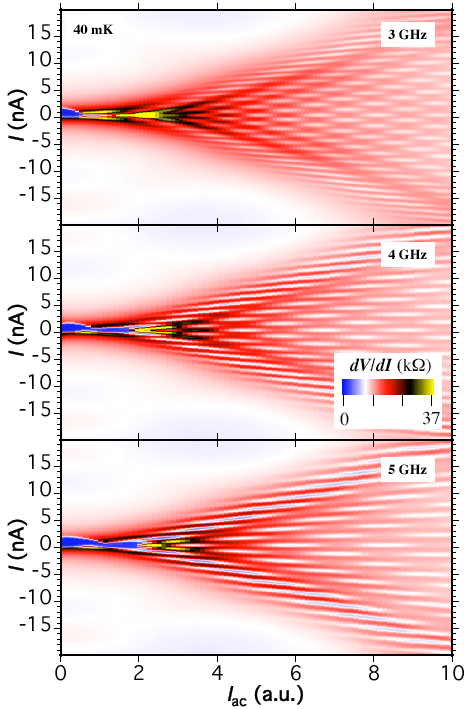}
\caption{(Color online) RF response in the off-state
at different microwave frequencies.
$dI/dV$-maps versus 
$I_{\rm sd}$ and $I_{\rm RF}$ are displayed for 40 mK. }
\label{figA3}
\end{center}
\end{figure}

The frequency dependence of the pattern of non-dissipative lobes
is present in the differential resistance maps of Figs.~\ref{figA2} and~\ref{figA3} 
for the on-state and off-state, respectively.

\end{document}